# BLADERUNNER

## Rapid Countermeasure for Synthetic (AI-Generated) StyleGAN Faces

Adam Dorian Wong
MIT Lincoln Laboratory
Group 52

01 September 2022

| | |
|---|---|
| MIT/LL: | adam.wong[at]ll.mit.edu |
| Gmail: | |
| | |
| Twitter: | @MalwreMorghulis |
| OTX: | MalwareMorghulis |
| | |
| GitHub (Personal): | https://github.com/MalwareMorghulis |
| GitHub (MIT/LL): | https://github.com/mit-ll/BLADERUNNER (DOI: 10.5281/zendo.7186014) |

This material is based upon work supported by the Department of the Air Force under Air Force Contract No. FA8702-15-D-0001. Any opinions, findings, conclusions or recommendations expressed in this material are those of the author(s) and do not necessarily reflect the views of the Department of the Air Force.

# Abstract

StyleGAN is NVIDIA's open-sourced TensorFlow implementation. It has revolutionized high quality facial image generation. However, this democratization of Artificial Intelligence / Machine Learning (AI/ML) algorithms has enabled hostile threat actors to establish cyber personas or sock-puppet accounts in social media platforms. These ultra-realistic synthetic faces. This report surveys the relevance of AI/ML with respect to Cyber & Information Operations. The proliferation of AI/ML algorithms has led to a rise in DeepFakes and inauthentic social media accounts. Threats are analyzed within the Strategic and Operational Environments. Existing methods of identifying synthetic faces exists, but they rely on human beings to visually scrutinize each photo for inconsistencies. However, through use of DLIBs' 68-landmark pre-trained file, it is possible to analyze and detect synthetic faces by exploiting repetitive behaviors in StyleGAN images. Project Blade Runner encompasses two scripts necessary to counter StyleGAN images. Through PapersPlease.py acting as the analyzer, it is possible to derive indicators-of-attack (IOA) from scraped image samples. These IOAs can be fed back into among_us.py acting as the detector to identify synthetic faces from live operational samples. The opensource copy of Blade Runner may lack additional unit tests and some functionality, but the open-source copy is a redacted version, far leaner, better optimized, and a proof-of-concept for the information security community. The desired end-state will be to incrementally add automation to stay on-par with its closed-source predecessor.

Going forward PapersPlease, papersplease.py, or papers_please.py naming schema will be used interchangeably, similar to Blade Runner and Among Us (AmongUs, among_us.py, or amongus.py).

Adam D. Wong
@MalwareMorghulis

# Introduction

Artificial Intelligence (AI) & Machine Learning (ML) are increasing areas of concern in the realm of cybersecurity. Adversaries are exploiting bleeding-edge technologies to enable cyber and information operations. Threat actors are leveraging DeepFakes and synthetic facial imagery to manipulate others. Democratization of AI/ML-based technologies poses a significant and continual cyber risk to national security. CNN reported that OpenAI refused to release their AI out of concern for abuse [1]. Open-source reporting suggests that proliferation of AI-generated imagery has been a key enabler for espionage, trolling, and harassment [2] [3].

DeepFakes remain a significant threat through projection of misinformation (misleading) and disinformation (deception) campaigns. However, one specific type of fake comes in the form of AI-generated synthetic facial images. Issues in the Strategic Environment (SE) derive from geo-political propriety (or lack thereof). International law has not kept up with advances in technology. For example, the Tallinn Manual has not yet addressed advances in AI/ML and its dangerous potential in war. The technology has not been adequately regulated either. In Operational Environments, AI/ML technologies are actively exploited by threat actors seeking to sew discord, maliciously influence others, or engage in social-engineering activities. People are prone to trust DeepFakes because they are getting more sophisticated in production and possibly based on cognitive biases. Social media personas leverage these synthetic photos which is a next-generation alternative to stolen real-photographs or generic stock images.

Blade Runner leverages pre-trained ML-predictor files to detect StyleGAN images through exploitable repetitive behaviors and Indicators-of-Attack (IOAs). Future iterations of open-source Blade Runner will automate certain tasks to stay on-par with its closed-source counterpart.

# Artificial Intelligence

In 2016, Microsoft created an AI chatbot: Tay on Twitter, to conduct an experiment in learning "conversations". The premise leveraged the idea that with more interaction (more data), AI would act more human through learning. However, Twitter users disrupted Tay-AI's learning and radicalize the chatbot [4]. Needless to say, AI learning can easily be disrupted by bad actors.

DeepFake is a portmanteau of "Deep Learning" [ML-technique] and "Fake [News]". The term was first observed in 2017 via a Reddit user account called: u/deepfakes [5]. This Redditor grafted celebrity faces onto pornographic media, but their account now lies devoid of any content [6]. This same technology has been used in Hollywood to posthumously revive beloved characters such as with: "Grand Moff Tarkin" and "Princess Leia" in *Star Wars: Rogue One* [7]. However, in the hands of threat actors, the technology has been abused to misinform or deceive audiences, degrade public-image, blackmail or embarrass geopolitical leaders, socially-engineer others, or sew discord in already chaotic environments [8] [9]. It's been used in memes where Hollywood actor Nicholas Cage's likeness is superimposed onto Harrison Ford's character in *Indiana Jones: Raiders of the Lost Ark* [10]. Nevertheless, the computer science areas of AI/ML are exponentially advancing. AI-assisted media manipulation will be a contested issue between freedom-of-research and misuse by hostile actors.

The capabilities have since been cyclically advanced and refined. Due to this nature of continuous improvement, technology for synthesizing DeepFakes has become common in industry. The technology has been democratized by different research entities and thus more openly-available to the public [11]. Academic research at leading technical universities has compelled the nation to expand in the AI/ML space [12] [13]. In bygone eras, this was called

"*doctoring*" media and "*photoshopping*" (based on Adobe products). Today, this they are known as *DeepFakes*.

## About StyleGAN

NVIDIA is a well-respected company making advances in the Graphical Processing Unit (GPU) and graphics card industry. Their department of research heavily operates in AI/ML-space. Their research in AI/ML has facilitated unparallel advances in computer-generated graphics. It must be acknowledged that StyleGAN allows ultra-realistic synthetic faces at 1024x1024 resolution.

***Generative Adversarial Networks (GANs)***. DeepFakes rely on multiple algorithms to synthesize images. Some images are manipulated through encoding processes to overlay latent images. However, some use ML-algorithms to intelligently overlap photos. Early forms of DeepFakes. It should be known that StyleGAN itself is not a malicious tool, nor was the research itself malicious. The application of StyleGAN by hostile actors is what makes the tool dangerous. Essentially, GANs rely on generators and discriminators to create these ultra-realistic images.

**Evolution of StyleGAN**:

- *2018*
    - NVIDIA proposed StyleGAN as their "official TensorFlow implementation" which used the Flickr-Faces-High-Quality (FFHQ) dataset to train their neural network [14] [15]
- *2019*
    - NVIDIA implemented improvements to the existing StyleGAN to enhance existing capabilities to generate synthetic imagery as StyleGAN 2 [16].
- *2021*
    - StyleGAN 3 acknowledges weaknesses of StyleGAN2 where certain features are placed based on fixed coordinates [17].
    - StyleGAN-NADA aims to use text to morph images from one graphic to another with limited training (such as Nvidia's example of a dog shifting to *The Joker*) [18].

# Proliferation of AI-Generated Images

*Availability of Tools*. Technology necessary to generated synthetic or grafted media has become easily-accessible to private individuals. For synthetic static imagery, two vendors are the most common: ThisPersonDoesNotExist (TPDNE) and GeneratedPhotos (GP). Dr. Lyu of University of Buffalo asserts that defense needs to evolve to counter AI/ML-counterfeiting advances [19]. For example: OpenAI [DALLE2](), [TensorFlow](), [GoogleAI](), and [PyTorch]() are openly available to the populace to use, test, or develop with. This very openness poses a risk of *information disruption* and it's arguable that one could equate this to concept of "script-kiddies" having access to DDoS tools. DDoS tools alone are stress testers for networks. However, simplicity of the tool and ease-of-access make them disruptive to enterprise networks. Open-source availability of AI/ML algorithms may be the same. However, societal benefits of AI/ML could very well outweigh risks of exploitation by bad actors.

> [*ThisPersonDoesNotExist*](). ThisPersonDoesNotExist (TPDNE) went viral in 2019 and implemented NVIDIA's StyleGAN solution. TPDNE was founded by Phillip Wang in demonstrating the power, capabilities, and looming threat for misuse of AI [20] [20]. Alternative, *DoesNotExist websites have been founded as well to demonstrate the power of StyleGAN and AI-generated imagery. TPDNE generates photos upon refresh. Currently, there is little user-customization and photos are generated at 1024x1024 resolution. The generation is sometimes imperfect.

> [*GeneratedPhotos*](). GeneratedPhotos (GP) gives user the opportunity to customize images based on background, faces, age, gender. It also provides a measure to buy or subscribe for bulk downloads of synthetic imagery. The freemium version of the tool generates photos at 512x512 resolution and for personal-use only.

Right, wrong, or indifferent – the democratization of AI/ML algorithms has led to a proliferation of AI-Generated Images misused by threat actors.

# Current Threats to the Strategic Environment

*Cyber Geneva Convention*. At the RSA Conference 2017, Brad Smith of Microsoft proposed the necessity of a "Cyber Geneva Convention" to establish precedent for conduct of hostilities in the Fifth Domain of Warfighting (Cyber) [21] [22]. Cyberwarfare is still a gray-zone in terms of the engagements between competing-powers. As the world enters an era of great-power competition between near-peer and peer adversaries, the likelihood of escalation from digital to kinetic warfighting increases. Given global trade and infrastructure heavily rely on cyber-physical systems, the cascading effects of crippled enterprises will lead to unintended (or intended) harm inflicted on civilians.

*Legal Complications*. As this report is being written, the Tallinn Manual 2.0 remains the current standard for International Customary Law with respect to cyber operations. Tallinn Manual 2.0 does not account for issues concerning AI/ML algorithms developed or deployed by nations. The "Black-Letter Rules" are bolded text yield the codified standards in policy and *jus ad bellum* [23]. Tallinn Manual 3.0 is currently being drafted by NATO Cooperative Cyber Defence Centre of Excellence (CCDCOE) [24]. At the time of this report, a recommendation was submitted by to Tallinn Manual workgroup to include a "Black-Letter Rule" for AI-algorithms.

One possibility is to apply export-controls onto AI/ML algorithms, but that would be a ceremonial response than practical. The European Union (EU) has attempted to regulate AI [25] [26]. This attempt of regulation of AI will not be without controversy.

*Information Warfare*. Cyber Operations focuses maneuver in and through networks and use of computers to provide effects upon a network [27]. Information Operations (IO) focuses on "integration of *influence*, *deception*, *corruption*, *usurping*", and changing perceptions of people by way of social media or another communication platform [28]. Rob M. Lee suggests intelligence activities belong somewhere between Defensive Cyber Operations – Response Actions and Offensive Security [29]. Doctrinally, these are distinct types of operations because the desired end-states are different. However, the mindset should shift to consider not two independent spectrums, but one unified plane. Operations are enabled by cyber and weaponization of information.

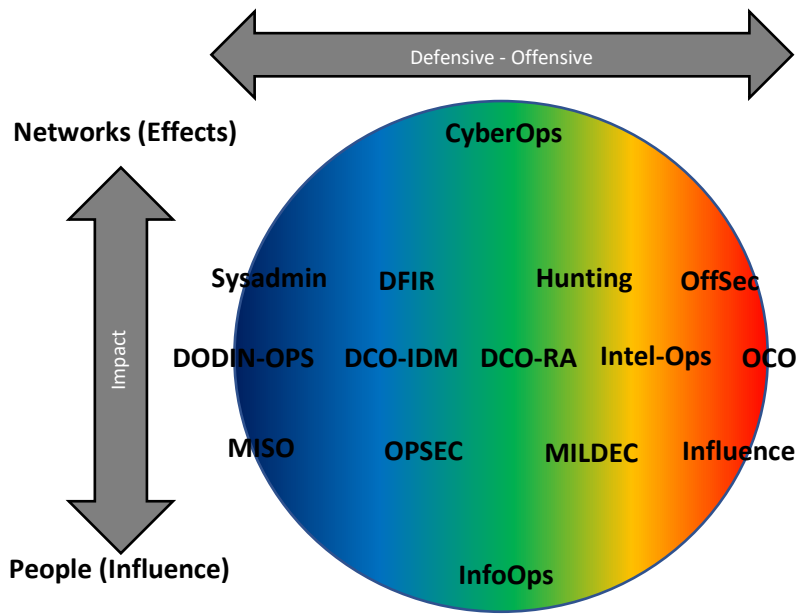

Figure 1: Proposed plane of Information Warfare

## Current Threats to the Operating Environment

*Social Media*. The introduction of social media has facilitated rapid communications across the globe. Most social media platforms rely on a combination of: username, email, phone number, date of birth, or photo to establish a digital persona. The problem is that platforms do not have adequate tools necessary to detect hostile actors leveraging AI-generated images. Controversy surrounding Twitter suggests that bots are a significant problem in social media [30]. By extension a subset of inauthentic accounts using synthetic faces.

*Cyber Operations*. StyleGAN has been a worry of the cybersecurity community for several years. Open-Source Intelligence (OSINT) Researcher and former SANS Institute Instructor, Micah Hoffman, identified the StyleGAN images as a photographic component for building cyber personas, commonly known as "sock-puppet" accounts in his SANS SEC487: OSINT Gathering & Analysis class [31]. Researchers at the Atlantic Council's Digital Forensic Research Lab (DFR Lab) noted issues as well [32]. A major concern is that these easily-created synthetic faces are being operationalized *en masse* for spear-phishing or illicit criminal activities.

*Information Operations*. The addition of AI-generated photos can be observed temporally. In Fall of 2019, Graphika investigated influence operations against Hong Kong by a threat actor dubbed "*Spamouflage Dragon*" (Graphika moniker) and Graphika found that influence operations on YouTube and Facebook leveraged stock photos [33]. This tactic would later evolve to incorporate StyleGAN images towards the end of 2019. Thereafter, a joint investigation by Graphika and Atlantic Council's DFRLab suggested that the majority of personas leveraged AI-generated photos [34]. It is perceived that Spamouflage Dragon later shifted operations to target the US General Election 2020, as well as agitate COVID-19 divisions [35]. In 2022, Graphika continued to monitor Spamouflage Dragon and found that the actor began engaging real-world (authentic) users with influence materials or propaganda and exited the established echo-chamber [36]. This change over the past several years indicate a shift of priority for capabilities, infrastructure, targeting, and desired impact.

Based on these facts, defenders are reminded that information and cyber operations are driven by humans behind the keyboard and therefore stratified by motivations or desires.

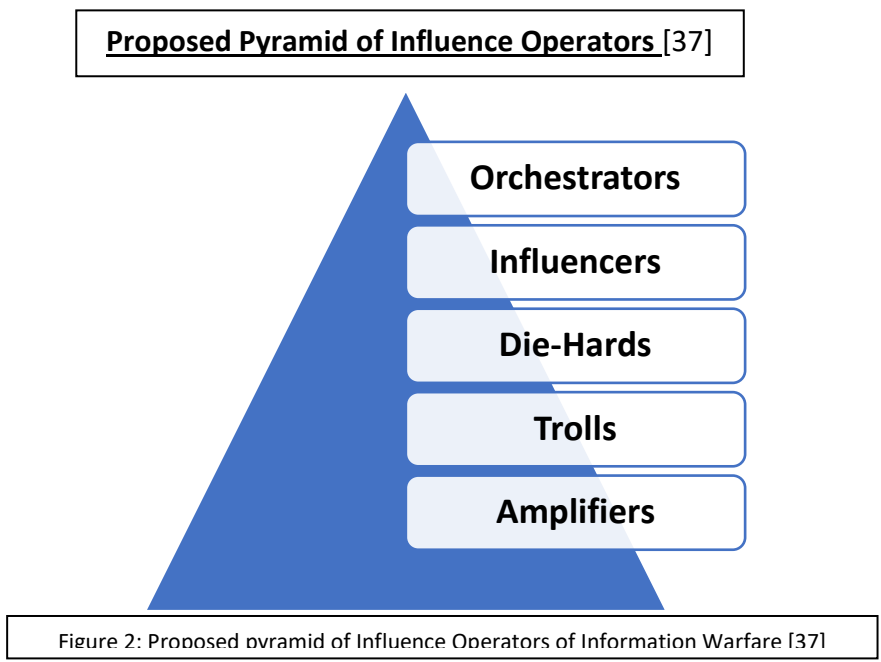

Figure 2: Proposed pyramid of Influence Operators of Information Warfare [37]

**Proposed Categories of Influence Actors** [37]

- **Orchestrator (*Manipulators*)**
    - Real hostile operator (social-engineering).
    - **Desire of Engagement**: Careful ~1:1 interaction to shape thinking or groom behavior.
- **Influencers (*Cult Leaders*)**
    - Real account.
    - **Desire of Popularity**: High follower count pushes message for them.
- **Die-Hards (*Fanatics*)**
    - Real people (sedated by propaganda).
    - **Desire of Dominance**: Engage in fights, build echo chambers, be-the-heroes.
- **Trolls (*Dividers & Disruptors*)**
    - Spambots or real people w/ destructive nature.
    - **Desire of Discord**: Engage in fights to sew resentment.
- **Amplifiers (*Spray n' Pray Msg.*)**
    - Spambots.
    - **Desire of Diffusion**: Botnet management & widest net cast.

*Marketing*. Surprisingly, recruiters are starting to use StyleGAN images for catfishing, social-engineering, or constructing sock-puppet accounts for head-hunting clients. Perceived intents of these inauthentic accounts could be innocent or malicious [38] [39]. By applying degrees-of-separation through counterfeit user profiles, there will continue to be a degradation of trust in communicative and social etiquettes between two persons. Not all threat actors are inherently criminals, agents of espionage, or warring parties. However, StyleGAN will continue to be abused outside of traditional inauspicious factions.

Threat actors will continue to use any and all means to appear valid and establish a digital beachhead by maintaining cyber personas or sock-puppet accounts, while projecting influence, continue targeting, supporting campaigns, and applying effects in the digital operating environment.

## Hypothesis on Trust in DeepFakes

Although the focus of this report in on countering synthetic images leveraged by inauthentic personas, it is worth identifying reasons why AI-generated images will be arduous to combat from human-based visual analysis alone.

***Biological (Symmetry & Attractiveness)***. It is possible that cognitive bias derives from subconscious assessments of attractiveness from facial symmetry. This symmetry is found in most of the generated StyleGAN photos. It is suggested that facial attractiveness has influence over socialization [40]. By extension, it is likely this bias carries over to with inauthentic accounts because the accounts use ultra-realistic images.

***Psychological (Halo Effect)***. Cognitive biases are always present in the human condition. In the 1920s, Edward Thorndike identified attractiveness as a contributor to cognitive bias [41] [42]. The term was later coined by S.M. Harvey in 1938 [41] [43]. A YouTube Channel: "Practical

Psychology", notes that the Halo Effect is essentially [unfairly] "judging a book by its cover" [44]. It is believed that this cognitive bias has undue influence over a person's ability to decide whether a social media persona is trustworthy (strictly based on photos alone).

*StyleGAN*. From a technological standpoint, DeepFakes are becoming harder to distinguish from real media (audio, photographic, video). Specifically, StyleGAN images are expected to mirror this trend. With StyleGAN2, production of synthetic faces has become increasingly harder to detect and more likely to be trusted by viewers [45]. Images are expected to become higher resolution.

*Hypothesis*. The current hypothesis is that facial symmetry is more likely to be trusted compared to asymmetrical ones [46]. The thesis and study conducted at George Mason University suggested there is perceived attractiveness in symmetry, however trustworthiness scoring was not validated due to errors in implementation and narrow sampling [46]. It's important to note that the "Halo Effect" is still contested in modern-day studies due to skewed results based on low sampling [47]. With media, video games, and images becoming higher resolution and AI algorithms becoming smarter through ML training, countermeasures for DeepFakes will need to evolve to keep hostile actors in-check.

# Current Detection Methods

Current countermeasures leverage ML to counter these AI-generated images. This is not a new concept. For instance, Silvia Man leveraged large face data sets and a ML neural network to determine whether or not AI-generated [48]. Silvia used modestly large datasets of faces derived from *UTKFace* (real) and *GeneratedPhotos* (synthetic). The distinction being ML-based (core components) which leverage significant training vs. ML-enabled (pre-trained datasets) having most of the core components being generic scripting or simple data manipulation.

***Brady Bunch Method***. Different research groups have placed the photos of identical sizes into matrices similar to the "Brady-Bunch" introductions. SANS Institute outlined this in their urgent webcast warning about personas leveraging synthetic photos during the Second Invasion of Ukraine in early 2022 [49]. This method provides a means to overlay lines or boxes and relies on visual analysis and intersection of photographic eyes. Based on the way photos are generated by StyleGAN, it appears that certain facial features like mouth corners and eyes are situated in identical locations across images. Because computers can't tell or don't care for these lines drawn on photos, this analysis relies on a human being to make a determination. Not to say a method cannot be coded – it can, but will take additional logic and functions. One drawback to the Brady Bunch method is that it relies on other photos in a set to be simultaneously analyzed together.

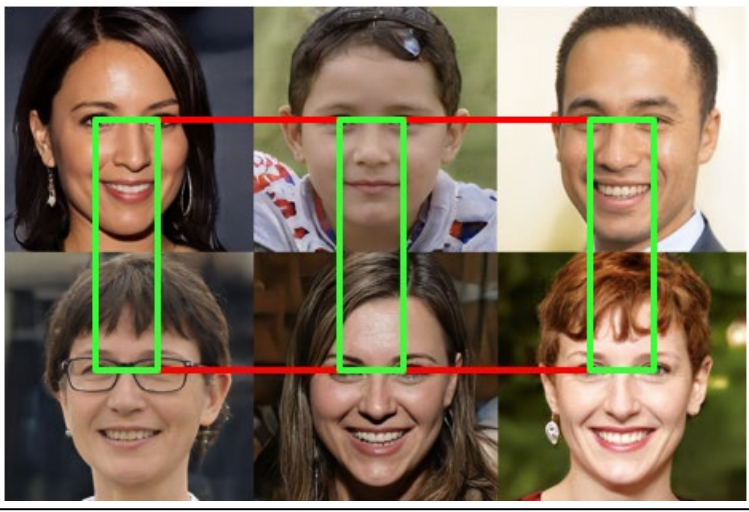

*Figure 3: Modified copy of observable characteristics in StyleGAN outlined by SANS Institute webcast* [49]
*(Image source: TPDNE)*

***Generation Quirks (TPDNE)***. StyleGAN has some observable oddities. More obvious issues are monstrosities in the imagery. It appears StyleGAN (at least the implementation by TPDNE) can only adequately generate one centered face. The input images may have included additional faces, but the GAN was unable separate or remove the faces. Another weakness of generation comes from mismatched apparel or wearable-accessories (*ie. glasses, earrings*) [50]. Other quirks include faces with hats that coincidentally turn into the hairline. These examples of are visible problems with StyleGAN and are visually-detectable fakes.

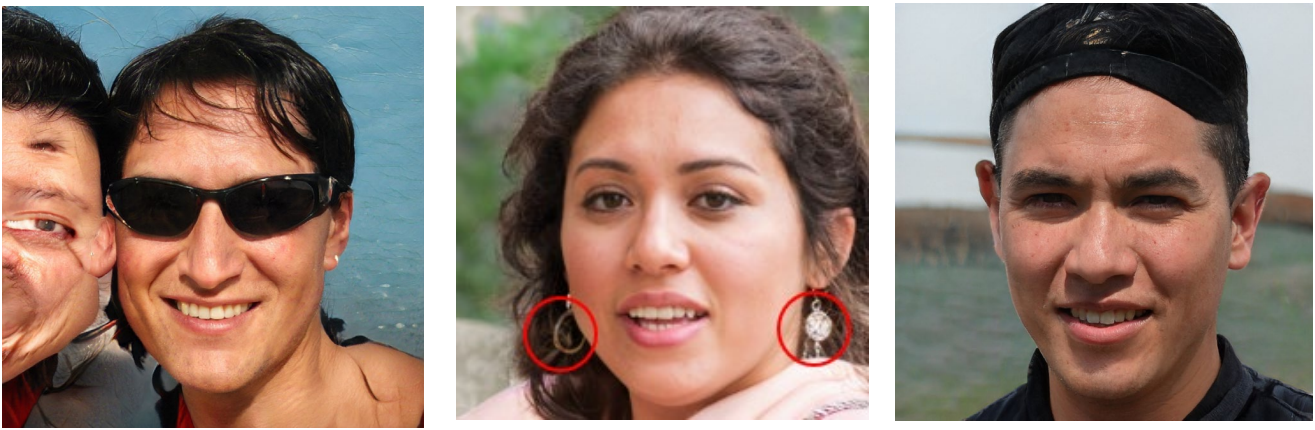

*Figure 4: Visual Quirks which from poor generation by StyleGAN. Left - Monstrosities, Middle - Mismatched Articles, Right - Headgear evolving into hairline (Image source: TPDNE)*

***Photo Overlay***. Overlaying photos is a technique used by image analysts for various functions. In 2019, a YouTuber named Atomic Shrimp, found that overlaying AI-generated photos would yield a generic face [51]. When AI-generated photos resurfaced in the InfoSec community, they were once again a topic of concern. In 2021, the Center of Information Resilience (CIR), suggested that eyes are generally fixed in location [52]. In 2022, a security researcher, Ben Nimmo

Adam D. Wong
@MalwareMorghulis                                       -13-

(Twitter @benimmo), attempted the same detection methodology and if defenders superimpose transparent copies of photos, then they would eventually yield mechanisms which eyes align [53]. One problem is that overlaying by itself does not necessarily act as a detection, but rather can support hypotheses from which to derive detection capabilities.

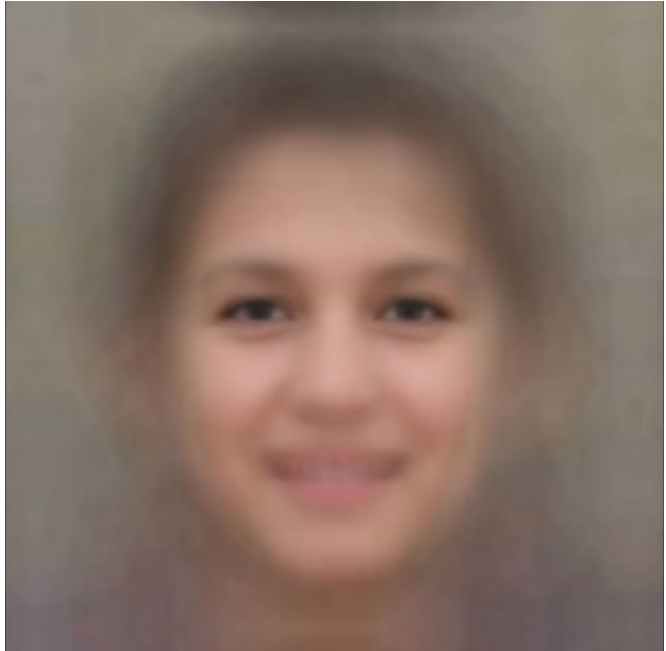

Figure 5: Photo Overlay Method using CombineZM by AtomicShrimp in 2019 (YouTube)

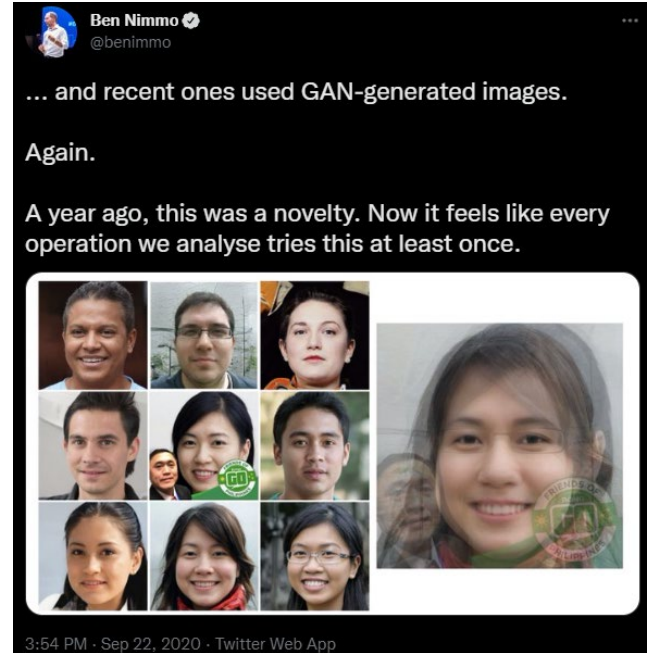

Figure 6: StyleGAN weaknesses observed by Ben Nimmo in 2020 (Twitter)

A problem with current detection methods is that it relies on end-users to visually scrutinize images for inconsistences such as blurry backgrounds, green clouds, mismatching jewelry, or asymmetrical facial features. Acknowledging without these prior efforts, Blade Runner would not be possible.

# Proposed Detection Method: Blade Runner

Many of the current methods for detecting AI-enabled fakes or DeepFakes rely on AI-countering-AI. SOC analysts can script solutions as stop-gap measures, but AI/ML are getting more advanced by-day. A major problem is SOC analysts may not have the in-house expertise in AI/ML DevOps (MLOps). SOC analysts may not have that expertise in this arena unless self-studying or having a strong background in Computer Science. Secondly, some organizations may not have the funding or resources to run large cloud-based systems or on-site supercomputers necessary to do large data-set analysis. Proliferation of AI/ML algorithms causes problems to increase exponentially especially when countering social-engineering where sock-puppets can be easily rotated, refreshed, reconstituted, or disposed. However, use of pre-trained ML libraries is a decent middle-ground where SOC analysts can simply script and the heavy-lifting of training AI/ML is already completed for them. Therefore, Blade Runner is necessary as a stop-gap measure to buy time for more permanent countermeasures.

For static images, simpler countermeasures are possible with pre-trained datasets. Typically, defenders are restrained by manpower, operational budgets, or vendor service costs. Using pre-trained libraries is an acceptable stop-gap solution which can be easier to implement or operationalize. Future detection methods may require significant use of AI/ML to counter-AI/ML and render human analysts obsolete. However, the immediate solution is to detect leverage pre-trained datasets and available tools to exploit existing problems with synthetic images.

***OpenCV and DLIB***. In 2017, Dr. Adrian Rosebrock noted that faces can be detected and outlined using a combination of Python, OpenCV, and DLIB [54]. Faces can be defined by the pre-trained landmark set by DLIB [54] [55]. DLIB's pretrained predictor file relies on 68-landmarks which can map to a face [56] [57]. With this pre-trained library, defenders can skip building their own ML-tool from scratch. Thus, saving time crafting a solution to counter synthetic images. However, DLIB requires images be converted to grayscale with OpenCV to work. Following this logic, it is possible to map the pre-trained outline onto an image of a person and thus build a detection against StyleGAN.

*Blade Runner Architecture*. This proposed detection method is composed of two components: PapersPlease (Analyzer) and AmongUs (Detector). PapersPlease becomes more accurate with more data scraped from publicly available websites like: [ThisPersonDoesNotExist](#) or [GeneratedPhotos](#). Blade Runner is intended to be a stop-gap solution to give defenders a chance to identify hostile actors and notify appropriate authorities.

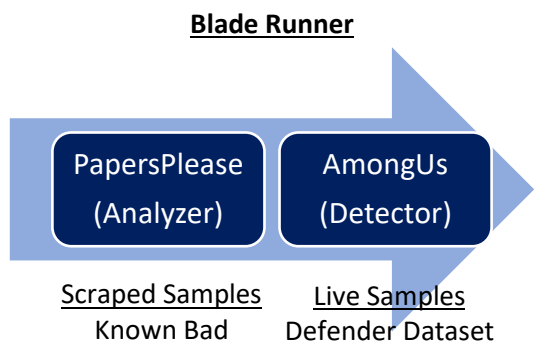

Figure 7: Blade Runner high-level architectural concept.

*PapersPlease (Analyzer)*. Based on mass analysis of data from PapersPlease, defenders can establish a goal-post or similarities across all known StyleGAN images (goal-posts being specific coordinate pairs for eyes). Defenders can derive coordinates for eyes for images across multiple resolutions. Samples analyzed by PapersPlease are large data-sets of synthetic faces. PapersPlease will scale images down from its original 1024x1024 resolution by Base-2 and Base-10 scales. It leverages the 68-landmark predictor in order to support the operations to determine center-points of eyes on an image of various resolutions. Albeit, some operations are still manually executed (at least in the open-source copy) and future work requires more automation. After deriving averages across different images and visual-testing, defenders can ascertain the center-of-eye coordinates. From here, data can be fed back into AmongUs. The premise is that with more StyleGAN image samples, the more accurate the coordinate-analysis will be.

*AmongUs (Detector)*. This script relies on pre-trained landmarking. Similarly, AmongUs specifically uses the [68-landmark predictor](). Based on this mapping, defenders can build quadrilateral polygons around centers of each eye (using landmarks on left-side: 37, 38, 40, 41 or right-side: 43, 44, 46, 47). These landmarks are planted, recorded, enumerated in sequential order and consistent across location mapping. Using the known coordinate goal-posts from PapersPlease, AmongUs is intended to be launched against live samples of images (whether synthetic or real) captured by network defenders via production security appliances. Defenders can test StyleGAN images and examine whether the sample image has eye-boxes which overlap the point-of-interest (goal-posts) (as derived from PapersPlease). For point and eye-box overlaps yielding True-True (Left, Right), defenders can reasonably assess that the image is likely StyleGAN-generated. Additionally, if eyes are in consistent locations, then the distance between the two will be constant. Other features such as nose or mouth width or height may be useful for future testing. Again, the open-source copy of Blade Runner does not yield as many unit-tests (or overt decision-making print-statements) for various reasons including speed-of-release. However, metadata is recorded for each image for future upgrades or performing new unit tests on images, and can be easily patched.

*Detection Weaknesses*. It must be acknowledged that the detection is less reliable when matched against synthetic faces wearing sunglasses (which obscure eye-outlines), especially with varying levels of tint. Secondly, data derived from papers_please.py is a mass average, false-positives have been observed for StyleGAN images glancing far-left or far-right and have been able to bypass the eye-overlap test. It is possible to detect them with the same methods, albeit with modified goal-post coordinates. At 100x100 pixel images, the detection is far-less reliable in images with full-body shots or with faces that do not take up the center-mass of the image. The detector has been known to fail where faces can be detected by OpenCV, but DLIB is incapable of mapping landmarks against faces. It is also a concern if an actor alters an image. Simply put, an adversary could carve away or delete a row or column of pixels to change the composition. In such case, the detection will need to account for movement of facial landmarks. The tool also does not detect use of generic stock images or stolen photos.

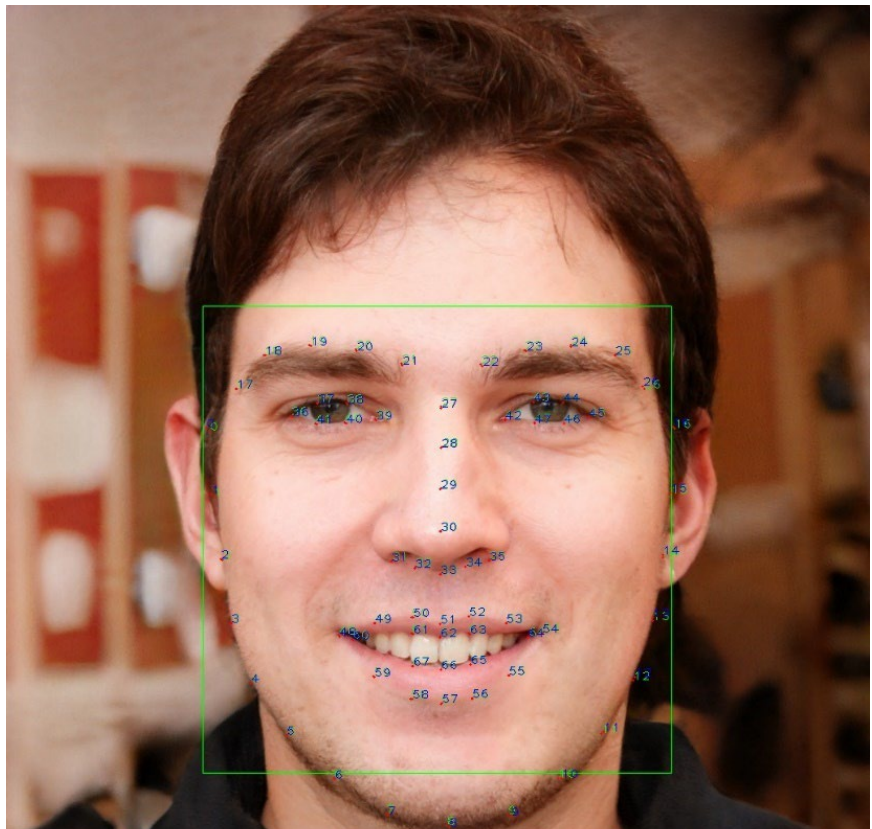

Figure 8: Applying Dr. Rosebrock's tracing technique against StyleGAN image [54]. (Image source: TPDNE)

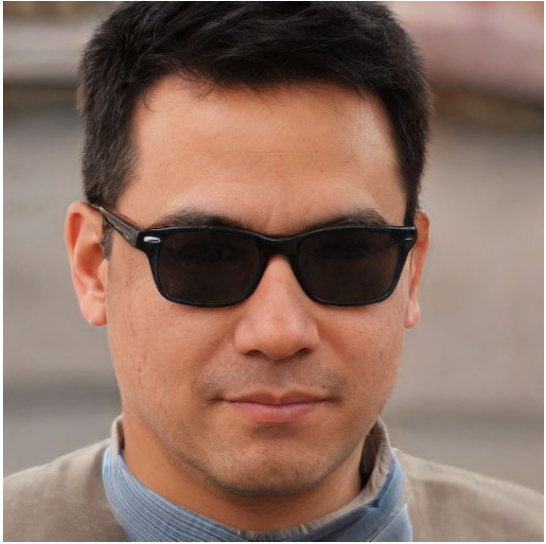 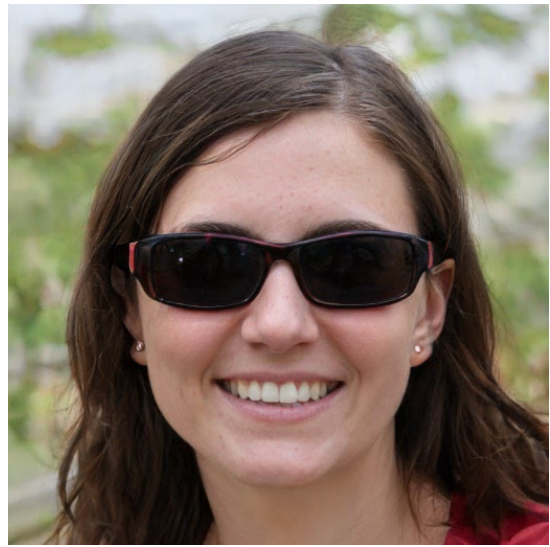

Figure 9: StyleGAN images with faces wearing sunglasses. (Image source: TPDNE)

# Exploitable Observations in StyleGAN2

- StyleGAN images are created through repetitive patterns, thus it can be exploited.
- The sum of the x-coordinates of each eye will approximately add up to the resolution width).
    - Without intensive code-review, the current hypothesis is that StyleGAN will divide square image into 3x equal-width vertical sections (depicted by "w").
    - StyleGAN likely hand-rails the two-cuts (red dotted-lines) to place eyes.
        - Meaning the AI-Generator may attempt calculate sites for eyes, plot the eyes first, and build a face around these hard-points and tries to do this by vertically dividing the image layout into approximate vertical-thirds.
    - The alternative analysis is that this is coincidence. However, this observable has been routinely seen across all scaled samples. This summation behavior is more apparent or prevalent when analyzing high number of image samples.
- The y-coordinates of the eyes (represented by blue dots) will always be just above* the horizontal half-way mark (yellow-line) of the image.
    - The resolution-height will be slightly more than the doubling the eye y-coordinate values.

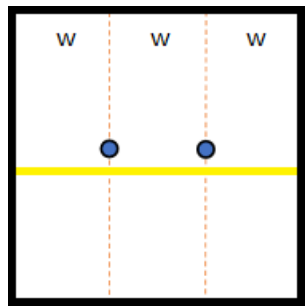

Figure 10: Generic outline for 1024x1024 image which StyleGAN may use to plot eyes.

- Due to the way the images are generated, the coordinates for each eye, will scale with the resolution (with roughly a 1–2-pixel margin-of-error).
    - The coordinates will naturally scale through:
        - Observed in Base-2 resolutions ($128... 256... 2^n$)
            *and*
        - Observed in multiples of Base-10 resolutions ($100... 200... 100n$)
    - This observation is currently based on limited testing for images with resolutions ≤ 1024x1024.
    - Therefore, we can reasonably predict where the next coordinates will be for up-scaled photos over 1024x1024.
    - If StyleGAN changes to incorporate higher resolution and larger images, then defenders have a chance to predict and detect it.
- The more data used in PapersPlease will lead to more accurate indicators-of-attack.

# Future Work

**Refactoring Blade Runner Project**. Several iterations of among_us.py have been crafted for various purposes and are active within production environments. However, it must be acknowledged that the open-source copy of Blade Runner is not as feature-rich as its closed-source counterpart – simply due to constraints of time and level-of-effort outside of normal business hours. The open-source copy is in a proof-of-concept and will be upgraded as time allows.

At the time of this report, papers_pllease.py uses imperfect CSV-write operations to record the eye-coordinates in StyleGAN images (extra headers written based on using Python PANDAS). Over time, other calculations will be added such as standard deviations. Hopefully, with larger datasets, defenders can ascertain distinctions between types of images (forward-looking, head-tilted, etc.). Defenders will need to mark samples similar to the Blade Runner (*ie*: A - Forward-facing, etc). The intent is to continuously upgrade or refactor Blade Runner and to better meet defender needs such as automating more mundane tasks: arithmetic operations, comparing, and contrasting StyleGAN image samples. Eye-locations (indicator-of-attack) were derived from manual execution of the analyzer and comparison with marked input image samples.

Papers_please.py requires additional testing to see if alternative types of synthetic faces have any significant deviations in facial features (such as face glancing left or right). *Does StyleGAN shift one of the eyes when glancing*? Papers_please.py does extract metadata and categorizes them based on pre-defined CSV files for future analysis.

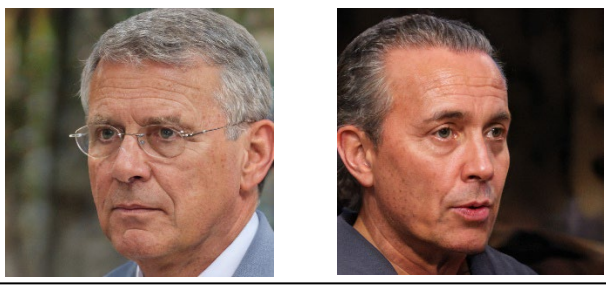
Figure 11: StyleGAN images glancing towards far left-right borders. (Image source: TPDNE)

Now, among_us.py still relies on hard-coded Boolean flags (for debugging purposes) and also will be optimized over-time. It is acknowledged that the code released with this report is still proof-of-concept, but worth sharing with the information security community for defensive purposes. Eventually, the open-source copy of Among Us will process multiple images instead of one per its proof-of-concept.

**Evolution of StyleGAN**. It is important to acknowledge a race condition that exists in cyber security between defenders and attackers. Each side operates in a contested domain and in a continuous and cyclical contestation for dominance. It is expected that StyleGAN will evolve due upgrade of the codebase as well as research by the originating vendor. Due to the advancements in AI/ML technology, it is guaranteed that successive iterations of StyleGAN will evolve beyond its known weaknesses [17]. It is very likely that future codebase forks or upgrades to StyleGAN will leverage alternative poses such as head-tilts or head locations or perhaps multiple faces per photo. Therefore, Blade Runner countermeasures will need to evolve to be able to counter personas or sock-puppets misusing synthetic imagery.

Adam D. Wong
@MalwareMorghulis                     -22-

# Conclusion

Democratization of AI/ML algorithms has led to advancements in image generation. AI/ML-assisted tools are the future of rapid solutioning for various problems: including lack of cloud or supercomputing hardware or even MLOps expertise. These tools have been misused by hostile actors to project influence or conduct attacks in the Fifth Domain. DeepFakes are a persistent threat to the Strategic and Operating Environments. StyleGAN has revolutionized synthetic imagery generation by providing ultra-realistic faces based on ML-training. However, threat actors are exploiting this technology to enable their campaigns in popular social media platforms. Current detection capabilities rely on visual analysis by human operators. Defenders need ML-enabled solutions to counter StyleGAN images. Without time-consuming processes of training ML-algorithms, it is possible to craft a rapid countermeasure to StyleGAN by using publicly available pre-trained models. Project Blade Runner (through PapersPlease and AmongUs) is the stopgap solution to detect and counter synthetic faces. The open-source version of the tool is still undergoing refinement to match similar (but not identical) capabilities of its closed-source version (without compromising proprietary information). It is expected that StyleGAN will evolve, thus defender capabilities must do the same.

# Acknowledgements

- Dr. Dennis Ross (05-52), *MIT Lincoln Laboratory* – Group Leader & Reviewer.

- Raul Harnasch (05-52) *MIT Lincoln Laboratory* – DevOps Mentor.

- Beau J. Guidry (05-52), *MIT Lincoln Laboratory* – Teammate, OSINT & DevOps.

- Steven Castellarin (11-10), *MIT Lincoln Laboratory* – SOC Mentor.

- Joshua Nadeau (12-10), *MIT Lincoln Laboratory* – DFIR Mentor.

- Alexander Rashash (12-10), *MIT Lincoln Laboratory* – CTAC.

- Micah Hoffman, *Spotlight-InfoSec, LLC*. – OSINT Instructor.

- Dr. Adrian Rosebrock, *PyImageSearch* – Coding tutorials in handling DLIB (Open-Source License).

- Davis E. King, *DLIB* – Provided DLIB 68-Landmarks Predictor & Pre-trained model.

- Project naming inspired by:
    - *Ladd Company* – Blade Runner, film (1982)
    - *3909 LLC* – Papers Please, video game (2013)
    - *InnerSloth LLC* – Among Us, video game (2018)

# References

[1]   R. Metz, "These People Do Not Exist. Why Websites Are Churning Out Fake Images of People (and Cats)," CNN, 28 February 2019. [Online]. Available: https://www.cnn.com/2019/02/28/tech/ai-fake-faces/index.html. [Accessed 22 August 2022].

[2]   "Experts: Spy Used AI-Generated Face to Connect with Targets," AP News, 13 June 2019. [Online]. Available: https://apnews.com/article/ap-top-news-artificial-intelligence-social-platforms-think-tanks-politics-bc2f19097a4c4fffaa00de6770b8a60d. [Accessed 22 August 2022].

[3]   K. Hill and J. White, "Designed to Deceive: Do These People Look Real to You?," New York Times, 21 November 2020. [Online]. Available: https://www.nytimes.com/interactive/2020/11/21/science/artificial-intelligence-fake-people-faces.html. [Accessed 22 August 2022].

[4]   J. Vincent, "Twitter Taught Microsoft's AI Chatbot to be a Racist ####### in Less Than a Day," TheVerge, 24 March 2016. [Online]. Available: https://www.theverge.com/2016/3/24/11297050/tay-microsoft-chatbot-racist. [Accessed 02 April 2022].

[5]   Reddit, "Reddit User 'u/deepfakes'," Reddit, 26 September 2017. [Online]. Available: https://www.reddit.com/user/deepfakes/. [Accessed 25 August 2022].

[6]   US Department of Homeland Security, "Increasing Threat of DeepFake Identities," 2022. [Online]. Available: https://www.dhs.gov/sites/default/files/publications/increasing_threats_of_deepfake_identities_0.pdf. [Accessed 25 August 2022].

[7]   S. Sarkar, "Rogue One Filmmakers Explain How They Digitally Recreated Two Characters," Polygon, 27 December 2016. [Online]. Available: https://www.polygon.com/2016/12/27/14092060/rogue-one-star-wars-grand-moff-tarkin-princess-leia. [Accessed 02 April 2022].

[8]   BuzzFeed, "You Won't Believe What Obama Says In This Video!," YouTube, 17 April 2018. [Online]. Available: https://www.youtube.com/watch?v=cQ54GDm1eL0. [Accessed 25 August 2022].

[9]   Business Standard, "Russia-Ukraine war: The Deceptive World of Deepfakes," YouTube, 20 March 2022. [Online]. Available: https://www.youtube.com/watch?v=y_4UsK4-RpQ. [Accessed 25 August 2022].

[10]  "Nicolas Cage is Indiana Jones in this Deepfake Video," CNET, 1 February 2018. [Online]. Available: https://www.cnet.com/culture/entertainment/machine-learning-algorithm-ai-nicolas-cage-movies-indiana-jones/. [Accessed 26 August 2022].

[11] Geeks For Geeks, "Top Open Source Projects Using Artificial Intelligence," Geeks For Geeks, 31 October 2020. [Online]. Available: https://www.geeksforgeeks.org/top-open-source-projects-using-artificial-intelligence/. [Accessed 25 August 2022].

[12] ThinkAI, "Top 10 Leading Universities in AI Research," ThinkAL, 09 August 2021. [Online]. Available: https://thinkml.ai/top-10-leading-universities-in-ai-data-science-research/. [Accessed 15 August 2022].

[13] US News, "Best Artificial Intelligence Programs in 2022," US News, 2022. [Online]. Available: https://www.usnews.com/best-graduate-schools/top-science-schools/artificial-intelligence-rankings. [Accessed 15 August 2022].

[14] T. Karras, S. Laine and T. Aila, "A Style-Based Generator Architecture for Generative Adversarial Networks," 29 March 2019. [Online]. Available: A Style-Based Generator Architecture for Generative Adversarial Networks. [Accessed 02 April 2022].

[15] NVLabs, "FFHQ-Dataset," GitHub, 2018. [Online]. Available: https://github.com/NVlabs/ffhq-dataset. [Accessed 26 August 2022].

[16] T. Karas, S. Laine, M. Aittala, J. Hellsten, J. Lehtinen and T. Aila, "Analyzing and Improving the Image Quality of StyleGAN," 23 March 2020. [Online]. Available: https://arxiv.org/pdf/1912.04958v2.pdf. [Accessed 02 April 2022].

[17] T. Karras, M. Aittala, S. Laine, E. Härkönen, J. Hellsten, J. Lehtinen and T. Aila, "Alias-Free Generative Adversarial Networks," 2021. [Online]. Available: https://nvlabs-fi-cdn.nvidia.com/stylegan3/stylegan3-paper.pdf. [Accessed 25 August 2022].

[18] R. Gal, O. Patashnik, H. Maron, A. Bermano, G. Chechik and D. Cohen-Or, "StyleGAN-NADA: CLIP-Guided Domain Adaptation of Image Generators," 16 December 2021. [Online]. Available: https://arxiv.org/pdf/2108.00946.pdf. [Accessed 25 August 2022].

[19] R. Banham and S. Lyu, "AI-Generated Facial Images Promise New Markets--and New Risk," Dell, 26 July 2021. [Online]. Available: https://www.dell.com/en-us/perspectives/ai-generated-facial-images-promise-new-markets-and-new-risk/. [Accessed 26 August 2022].

[20] D. Paez, "'This Person Does Not Exist' Creator Reveals His Site's Creepy Origin Story," Inverse, 21 February 2019. [Online]. Available: https://www.inverse.com/article/53414-this-person-does-not-exist-creator-interview. [Accessed 02 April 2022].

[21] B. Smith, "The Need for a Digital Geneva Convention," Microsoft, 14 February 2017. [Online]. Available: https://blogs.microsoft.com/on-the-issues/2017/02/14/need-digital-geneva-convention/. [Accessed 15 August 2022].

[22] B. Smith, "RSA Conference 2017: Protecting and Defending against Cyberthreats in Uncertain Times," YouTube, 15 February 2017. [Online]. Available: https://www.youtube.com/watch?v=kP_yf_Uz4vc. [Accessed 15 August 2022].

[23] M. N. Schmitt, Tallinn Manual 2.0 on the International Law Applicable to Cyber Operations, Cambridge: Cambridge University Press, 2017.

[24] M. Schmitt, "CCDCOE to Host the Tallinn Manual 3.0 Process," NATO CCDCOE, 2022. [Online]. Available: https://ccdcoe.org/news/2020/ccdcoe-to-host-the-tallinn-manual-3-0-process/. [Accessed 26 August 2022].

[25] "Artificial Intelligence Act," European Union, April 2021. [Online]. Available: https://artificialintelligenceact.eu/. [Accessed 26 August 2022].

[26] J. Meltzer and A. Tielemans, "The European Union AI Act," May 2022. [Online]. Available: https://www.brookings.edu/wp-content/uploads/2022/05/FCAI-Policy-Brief_Final_060122.pdf. [Accessed 26 August 2022].

[27] US Joint Chiefs of Staff, "Joint Publication 3-12: Cyber Operations," 08 June 2018. [Online]. Available: https://www.jcs.mil/Portals/36/Documents/Doctrine/pubs/jp3_12.pdf. [Accessed 26 August 2022].

[28] US Joint Chiefs of Staff, "Joint Publication 3-13: Information Operations," 27 November 2012. [Online]. Available: https://www.jcs.mil/Portals/36/Documents/Doctrine/pubs/jp3_13.pdf. [Accessed 25 August 2022].

[29] R. M. Lee, "The Sliding Scale of Cyber Security," SANS Institute, 2015.

[30] R. McMillan, "Twitter Whistleblower Peiter Zatko Has Warned of Cyber Disasters for Decades," The Wall Street Journal, 24 August 2022. [Online]. Available: https://www.wsj.com/articles/twitter-whistleblower-peiter-zatko-has-warned-of-cyber-disasters-for-decades-11661361357. [Accessed 26 August 2022].

[31] M. Hoffman, SEC487: OSINT Gathering & Analysis, Bethesda: SANS Institute, 2018.

[32] DFRLab, "Inauthentic Instagram Accounts with Synthetic Faces Target Navalny Protests," Atlantic Council DFRLab, 28 January 2021. [Online]. Available: https://medium.com/dfrlab/inauthentic-instagram-accounts-with-synthetic-faces-target-navalny-protests-a6a516395e25. [Accessed 27 August 2022].

[33] B. Nimmo, C. S. Eib and L. Tamora, "Graphika Report: Spamouflage," 25 September 2019. [Online]. Available: https://public-assets.graphika.com/reports/graphika_report_spamouflage.pdf. [Accessed 01 February 2022].

[34] K. Johnson, B. Nimmo, S. E. C., L. Tamora, I. Smith, E. Buziashvili, A. Kann, K. Karan, E. P. d. L. Rosas, M. Rizzuto, C. François and I. Robertson, "#OperationFFS: Fake Face Swarm," December 2019. [Online]. Available: https://public-assets.graphika.com/reports/graphika_report_operation_ffs_fake_face_storm.pdf. [Accessed 01 February 2022].

[35] B. Nimmo, C. S. Eib, C. François and L. Ronzaud, "Spamouflage Goes to America," 12 August 2020. [Online]. Available: https://public-

assets.graphika.com/reports/graphika_report_spamouflage_goes_to_america.pdf. [Accessed 01 February 2022].

[36] B. Nimmo, I. Hubert and Y. Cheng, "Spamouflage Breakout," 04 February 2021. [Online]. Available: Ben Nimmo, Ira Hubert and Yang Cheng. [Accessed 28 February 2022].

[37] A. D. Wong, *Beaver Works Cyber Operations Module 1-2 OSINT,* Lexington: MIT/LL Beaver Works, 2022.

[38] A. Lusina, "Researchers Say LinkedIn is Overrun with Fake, AI-Generated Profiles," PetaPixel, 29 March 2022. [Online]. Available: https://petapixel.com/2022/03/29/researchers-discover-more-than-1000-ai-generated-linkedin-profiles/. [Accessed 26 August 2022].

[39] S. Bond, "That Smiling LinkedIn Profile Face Might be a Computer-Generated Fake," NPR, 27 March 2022. [Online]. Available: https://www.npr.org/2022/03/27/1088140809/fake-linkedin-profiles. [Accessed 26 August 2022].

[40] A. C. Little, B. C. Jones and L. M. DeBruine, "Facial Attractiveness: Evolutionary-Based Research," *Philosophical Transactions of the Royal Society B Biological Science,* vol. 366, no. 1571, p. 1638–1659, 2011.

[41] The Decision Lab, "The Halo Effect, Explained," The Decision Lab, [Online]. Available: https://thedecisionlab.com/biases/halo-effect. [Accessed 15 August 2022].

[42] E. L. Thorndike, "A Constant Error in Psychological Ratings," *Journal of Applied Psychology,* vol. 4, no. 1, p. 25–29, 1920.

[43] S. M. Harvey, "A Preliminary Investigation of the Interview," *British Journal of Psychology,* vol. 28, no. 3, p. 263–287, 1938.

[44] Practical Psychology, "The Halo Effect - How Attractiveness Flows using the Psychology of Attraction," YouTube, 22 January 2018. [Online]. Available: https://www.youtube.com/watch?v=2h6HeqO-U9c. [Accessed 15 August 2022].

[45] S. J. Nightingale and H. Farid, "AI-Synthesized Faces are Indistinguishable from Real Faces and More Trustworthy," *Proceedings of the National Academy of Sciences,* vol. 119, no. 8, p. 2120481119, 2021.

[46] L. Dupuis, "A Behavioral Study of Face Symmetry and Trustworthiness," 2019. [Online]. Available: http://mars.gmu.edu/bitstream/handle/1920/11797/Dupuis_thesis_2019.pdf?sequence=1&isAllowed=y#:~:text=A%20common%20contributor%20to%20ratings,symmetry%20in%20judgments%20of%20faces.. [Accessed 15 August 2022].

[47] L. Graves, D. Ellers, S. Troyer and S. Welch, "The Effects of Facial Symmetry on Perceived Attractiveness and Trustworthiness," 2019. [Online]. Available: http://people.uncw.edu/noeln/documents/Spring2019Facialsymmetry.pdf. [Accessed 15 August 2022].

[48] S. Man, "Using ML to Detect Fake Face Images Created by AI," Jayway Blog by DevoTeam, 06 March 2020. [Online]. Available: https://blog.jayway.com/2020/03/06/using-ml-to-detect-fake-face-images-created-by-ai/. [Accessed 18 August 2022].

[49] R. Lee, K. Holvoet, T. Conway and J. Williams, "Urgent Webcast: Russian Cyber Attack Escalation in Ukraine - What You Need To Know!," 25 February 2022. [Online]. Available: https://www.sans.org/webcasts/russian-cyber-attack-escalation-in-ukraine/. [Accessed 25 August 2022].

[50] B. Collins, "@OneUnderscore_," Twitter, 28 February 2022. [Online]. Available: https://twitter.com/ct_bergstrom/status/1498887434783784967?lang=en. [Accessed 26 August 2022].

[51] Atomic Shrimp, "The Average Face Of ThisPersonDoesNotExist.com," YouTube, 18 November 2019. [Online]. Available: https://www.youtube.com/watch?v=CrTRbsnfcgc. [Accessed 30 June 2022].

[52] B. Strick, "Revealed: Coordinated Attempt to Push Pro-China, Anti-Western Narratives on Social Media," 05 August 2021. [Online]. Available: https://www.info-res.org/post/revealed-coordinated-attempt-to-push-pro-china-anti-western-narratives-on-social-media. [Accessed 22 August 2022].

[53] B. Nimmo, Twitter, 28 February 2022. [Online]. Available: https://twitter.com/benimmo/status/1498406275490582530. [Accessed 22 August 2022].

[54] A. Rosebrock, "Facial landmarks with dlib, OpenCV, and Python," 03 April 2017. [Online]. Available: https://pyimagesearch.com/2017/04/03/facial-landmarks-dlib-opencv-python/. [Accessed 06 February 2022].

[55] D. E. King, "dlib C++ Library," 2022. [Online]. Available: http://dlib.net/files/shape_predictor_68_face_landmarks.dat.bz2. [Accessed 06 February 2022].

[56] Imperial College London, "Intelligent Behaviour Understanding Group (iBUG): Facial Point Annotations," Imperial College London, [Online]. Available: https://ibug.doc.ic.ac.uk/resources/facial-point-annotations/. [Accessed 28 February 2022].

[57] DLIB, "Shape Predictor 68-Face Landmarks," DLIB, [Online]. Available: http://dlib.net/files/shape_predictor_68_face_landmarks.dat.bz2.